\documentclass[%
reprint, 
superscriptaddress,
bibnotes,
amsmath,amssymb,
aps,
]{revtex4-1}

\usepackage{graphicx}
\usepackage{dcolumn}
\usepackage{bm}
\usepackage{multirow}
\usepackage{color}
\usepackage{comment}
\definecolor{blue2}{rgb}{0,0.6,1}

\begin{document}
\title{Enhancement of ablative Rayleigh-Taylor instability growth by thermal conduction suppression in a magnetic field}

\author{Kazuki Matsuo}
\email{matsuo-k@ile.osaka-u.ac.jp}
\affiliation{Institute of Laser Engineering, Osaka University, Suita,
   Osaka 565-0871, Japan}
   
\author{Takayoshi Sano}
\affiliation{Institute of Laser Engineering, Osaka University, Suita,
   Osaka 565-0871, Japan}
   
\author{Hideo Nagatomo}
\affiliation{Institute of Laser Engineering, Osaka University, Suita,
   Osaka 565-0871, Japan}
   
\author{Toshihiro Somekawa}
\affiliation{Institute of Laser Engineering, Osaka University, Suita,
   Osaka 565-0871, Japan}
\affiliation{Institute for Laser Technology, 1-8-4 Utsubo-honmachi, Nishi-ku Osaka, Osaka, 550-0004, Japan.}
   
\author{King Fai Farley Law}
\affiliation{Institute of Laser Engineering, Osaka University, Suita,
   Osaka 565-0871, Japan}
   
\author{Hiroki Morita}
\affiliation{Institute of Laser Engineering, Osaka University, Suita,
   Osaka 565-0871, Japan}
   
\author{Yasunobu Arikawa}
\affiliation{Institute of Laser Engineering, Osaka University, Suita,
   Osaka 565-0871, Japan}
   
\author{Shinsuke Fujioka}
\affiliation{Institute of Laser Engineering, Osaka University, Suita,
   Osaka 565-0871, Japan}


\date{\today}
             
\begin{abstract}
Ablative Rayleigh-Taylor instability growth was investigated to elucidate the fundamental physics of thermal conduction suppression in a magnetic field.
Experiments found that unstable modulation growth is faster in an external magnetic field.
This result was reproduced by a magnetohydrodynamic simulation based on a Braginskii model of electron thermal transport.
An external magnetic field reduces the electron thermal conduction across the magnetic field lines because the Larmor radius of the thermal electrons in the field is much shorter than the temperature scale length.
Thermal conduction suppression leads to spatially nonuniform pressure and reduced thermal ablative stabilization, which in turn increases the growth of ablative Rayleigh-Taylor instability.
\end{abstract}
\maketitle


Inertial confinement fusion (ICF) is created by imploding a spherical target to achieve high compression of the fuel and generate a high temperature hot spot to trigger ignition and maximize the thermonuclear energy gain.
While progress towards the ICF ignition is being made at many research facilities\cite{LePape2018, Regan2016}, the current central ignition scheme has not yet reached the ignition condition.
Hot spark mixing with the cold dense fuel hampers fusion ignition mainly due to the significant growth of Rayleigh-Taylor (RT) instabilities\cite{Zhou2017, Zhou2017a} during the compression.
Many current ICF research efforts are directed at understanding and controlling the growth of these asymmetries.

An alternate approach is to accept that perturbations are unavoidable in ICF experiments and instead reduce hot-spot cooling through the application of an external magnetic field.
Perkins \textit{et al.}\cite{Perkins2013} found that the application of a strong magnetic field to fusion targets relaxes the ignition requirements in two-dimensional (2D) magnetohydrodynamic (MHD) simulations. 
The optimum magnetic field enhanced the fusion yield by a factor of over 200. 
The non-uniformity of shape asymmetries increased with increasing field strength, giving an optimum applied field of around 50 $\mathrm{T}$. 
Walsh \textit{et al.}\cite{Walsh2019} also found that the ignition requirements relaxed when hot-spot cooling was reduced in 3D extended-MHD simulations. RT growth increases in a magnetic field due to the reduced thermal ablative stabilization.
Pre-magnetization of ICF implosions has the potential to enhance the fusion yield due to the reduction of hot-spot cooling. 
On the other hand, enhancement of perturbation growth due to the magnetic field is unavoidable.  Experimental verification of both phenomena will be the key to the success of this scheme.

Laser-driven magnetic field compression of ICF targets has been performed by Chang \textit{et al.}\cite{Chang2011} on the OMEGA laser facility with seed fields of 8 $\mathrm{T}$. 
Measurable increases in ion temperatures and neutron yields were detected, resulting from reduced electron heat conduction. 
However, because the seed magnetic field was not large enough to affect the perturbation growth, the unavoidable symmetry changes due to the applied magnetic field have not been measured.

In the current study, RT instability growth experiments were performed to demonstrate thermal conduction suppression due to a magnetic field. 
Thermal conduction suppression leads to spatially nonuniform pressure and a reduced thermal ablative stabilization, which in turn increases the growth of ablative RT instability.
Unstable modulation growth in an external magnetic field was strongly increased compared to the growth in the absence of the field.
It is essential for future experimental designs to obtain the extent to which the magnetic field affects the ablative RT instability and the conditions that the magnetic field can affect.

First, we will explain the theoretical background briefly. Assuming a single fluid and two-temperature plasma, the conservation of momentum and electron energy of the resistive MHD in an external magnetic field are described as:
\begin{eqnarray}
\rho\frac{\textrm{d}\mathbf{u}}{\textrm{d}t}&=&-\nabla(P_{\mathrm{th}}+\frac{\mathbf{B}^2}{8 \pi})+\frac{1}{4 \pi}(\mathbf{B} \cdot \nabla)\mathbf{B}, \label{eq:momentum}\\
\rho\frac{\textrm{d}\varepsilon_e}{\textrm{d}t}&=&-P_e\nabla \cdot \mathbf{u}-\nabla \cdot (\boldsymbol{\kappa} \cdot \nabla T_e)-Q_{ei}+S_L+S_r \label{eq:electron energy}.
\end{eqnarray}
Here, $\rho$ and $\mathbf{u}$ are the mass density and fluid velocity of the plasma, respectively, $P_e$ is the pressure of the electron, $Q_{ei}$ is the ion--electron relaxation, and $S_L$ and $S_r$ are the energy transferred from the laser and by radiation to the plasma, respectively.

The influence of the external magnetic field on the plasma manifests in two ways. First, the plasma motion is directly changed by the Lorentz force in Eq.\ref{eq:momentum}, though this effect is relatively small in our high-$\beta$ plasma, where the parameter $\beta$ is the ratio of the thermal and magnetic pressures:
\begin{equation}
    \beta =4.0 \left(\frac{B}{100 \mathrm{T}} \right)^{-2} \left(\frac{T_e}{100 \mathrm{eV}} \right) \left(\frac{n_e}{10^{21} \mathrm{cm}^{-3}} \right).
\end{equation}
Hydrodynamic instability growth is reduced by the restitution force of the magnetic field bent by non-uniform plasma flow in a low-$\beta$ plasma and for an external magnetic field of critical strength \cite{Stone2007, Sano2013}, whereas the thermal pressure is always larger than the magnetic pressure in this study (i.e., $\beta >> 1$).

Second, the electron energy increment in Eq. \ref{eq:electron energy} is modified by the magnetic field through the thermal conductivity, where $\boldsymbol{\kappa}$ is the thermal conductivity tensor. Assuming 2D transport, $\boldsymbol{\kappa}$ is described as:
\begin{eqnarray}
\boldsymbol{\kappa} \cdot \nabla T_e&=&\kappa_{\parallel} \nabla_{\parallel}T_e+\kappa_{\bot} \nabla_{\bot}T_e + \kappa_{\wedge} \nabla T_e,\\
\kappa_{\parallel}&=&\gamma_0(\frac{n_e T_e \tau_e}{m_e}) = \gamma_0\kappa_0,\\
\kappa_{\bot}&=&\kappa_0 \frac{\gamma'_1 \chi^2+\gamma'_0} {\bigtriangleup} ,\\
\kappa_{\wedge}&=&\kappa_0 \frac{\chi \left( \gamma''_1 \chi^2+\gamma''_0 \right)}{\bigtriangleup},
\end{eqnarray}
where $\chi$ is the Hall parameter and $\bigtriangleup=\chi^4+\delta_1 \chi^2+\delta_0$. $\gamma'_1$, $\gamma'_0$, $\delta_1$, and $\delta_0$ are the Braginskii coefficients\cite{Braginskii1965}, whose values vary with the magnetic field. 

The Hall parameter is the product of the electron gyrofrequency ($\omega_\textrm{c}$) and electron-ion collision time ($\tau_\textrm{ei}$):
\begin{equation}
    \omega_\textrm{c} \tau_\textrm{ei} =1.3 \left(\frac{B}{100 \mathrm{T}} \right) \left(\frac{T_e}{100 \mathrm{eV}} \right)^{\frac{3}{2}} \left(\frac{n_e}{10^{21} \mathrm{cm}^{-3}} \right)^{-1}.
\end{equation}
For a 300-eV polystyrene plasma, 200-T external magnetic field, and critical density for 351-nm laser beams, the Hall parameter is close to unity. 
When the Hall parameter is non-zero, the external magnetic field reduces the electron thermal conduction across the magnetic field lines because the Larmor radius of the thermal electrons in the magnetic field is much shorter than the mean-free-path of the thermal electrons. We found that the thermal conduction suppression due to an external magnetic field affects RT instability growth even in our high-$\beta$ plasma. 

A basic experiment with a simple geometry was performed with a spatially uniform strong magnetic field generated by a pair of laser-driven capacitor coil targets.
Three GEKKO-XII laser beams were used for each capacitor coil target to generate the magnetic field.
The wavelength, pulse shape, pulse duration, and energy of the GEKKO-XII beams were 1.053 $\mu$m, Gaussian, 1.2 ns full-width at half-maximum, and 700 $\pm$ 20 J per beam.
The strength of the magnetic field generated with the capacitor-coil target was measured on the GEKKO-XII, LULI2000, Shengguang-II, and OMEGA-EP laser facilities \cite{Law2016, Fujioka2013, Zhu2015, Santos2015b, Gao2016}.
215 $\pm$ 21 $\mathrm{T}$ magnetic fields were obtained in a previous experiment\cite{Matsuo2016} with the same configuration.

\begin{figure}[ht]
\begin{center}
 \includegraphics[width=80mm]{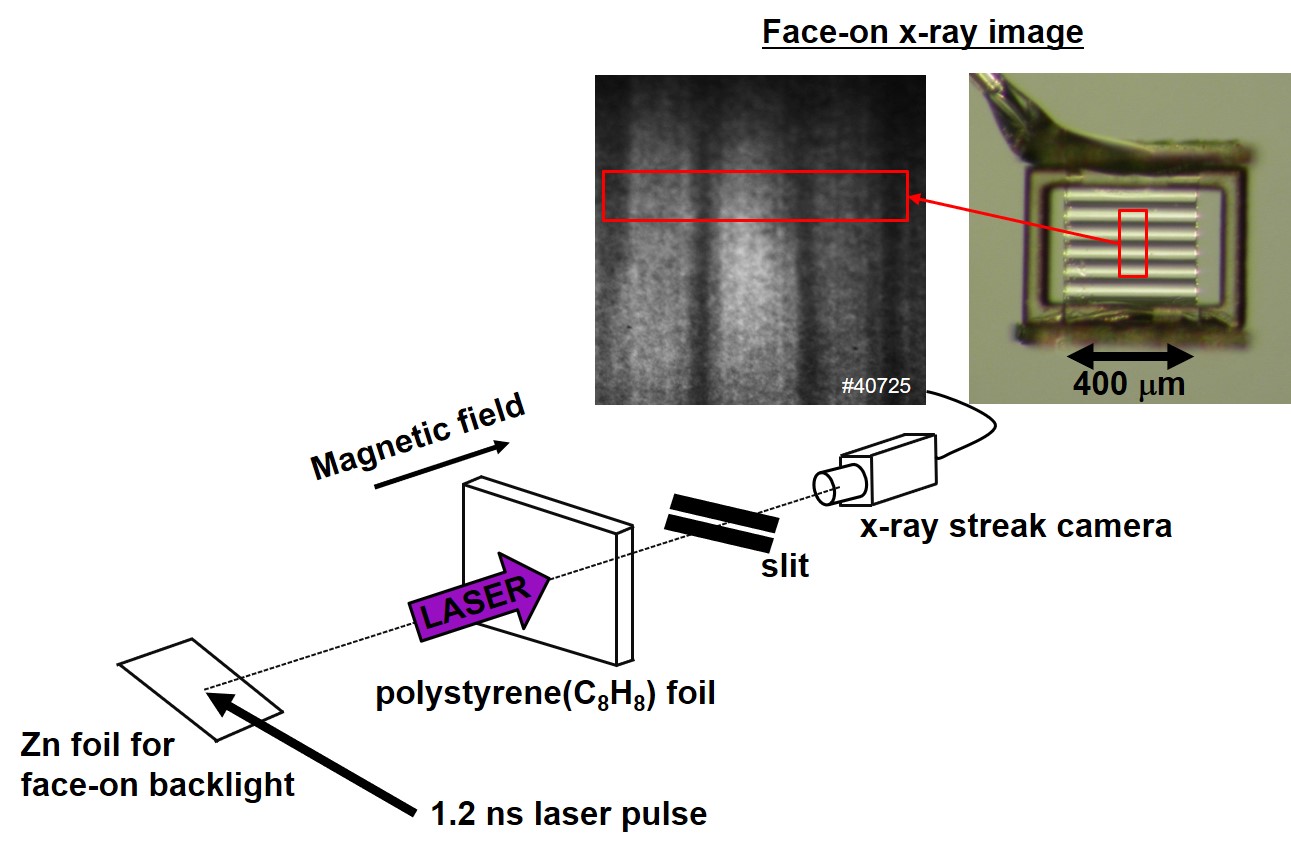}
\end{center}
 \caption{Experimental setup. Face-on x-ray backlighting was used to measure the temporal evolution of the areal density modulations.} \label{fig:Figure1}
\end{figure}

Figure \ref{fig:Figure1} shows the experimental layout with a photograph of the target. A 16-$\mu$m-thick polystyrene (C$_8$H$_8$) foil was irradiated with laser beams midway between the two coils. The external magnetic field is in the same direction as the incoming laser beam.
The width of the polystyrene foil was 400 $\mu$m.
Initial sinusoidal perturbations for 30, 60, and 100-$\mu$m wavelengths ($\lambda$) with an initial amplitude 1.0 $\pm$ 0.1 $\mu$m were incident on the front side of the planar polystyrene foils.
15-$\mu$m tantrum plates were placed at the bottom and top of the polystyrene foil to prevent the polystyrene foil from being preheated by x-rays generated at the capacitor parts.
Three 351-nm beams of the GEKKO-XII laser were used to drive the polystyrene foil at an intensity of (2.5 $\pm$ 0.1) $\times$ $10^{14}$ $\mathrm{W/cm^2}$. 

The polystyrene foil with the incident perturbation is accelerated by the drive laser and is measured by backlighting x-rays emitted from a zinc foil that is irradiated by a separate laser.
The laser-produced zinc plasmas emit relatively broad $L$-shell x-rays centered at 1.5 keV \cite{Azechi2007}. 5.5-$\mu$m-thick Al ($K$ absorption edge at 1.56 keV) and 25-$\mu$m-thick Be foils were placed in front of an x-ray streak camera (XSC) for x-ray filtering. 

The intensity distribution of the x-rays transmitted through the target is imaged on the XSC.
The areal-density perturbation is thus recorded as the contrast of the x-ray intensity distribution,
a technique known as "face-on x-ray backlighting" \cite{Azechi2007, Sakaiya2002}.
Face-on x-ray backlighting was used to measure the temporal evolution of the areal density modulations ($\Delta \rho a$) amplified by the hydrodynamic instability from the x-ray intensity ratio between the peaks ($I_\textrm{peak}$) and valleys ($I_\textrm{valley}$) of an image as $\Delta \rho a = \ln(I_\textrm{peak}/I_\textrm{valley})/2 \mu$, where $\mu$ = 607 g/cm$^2$ is the mass absorption rate of polystyrene for the x-rays. The spatial and temporal resolutions of the x-ray imaging system were measured to be 13 $\mu$m and 130 ps, respectively.

\begin{figure}[ht]
\begin{center}
 \includegraphics[width=75mm]{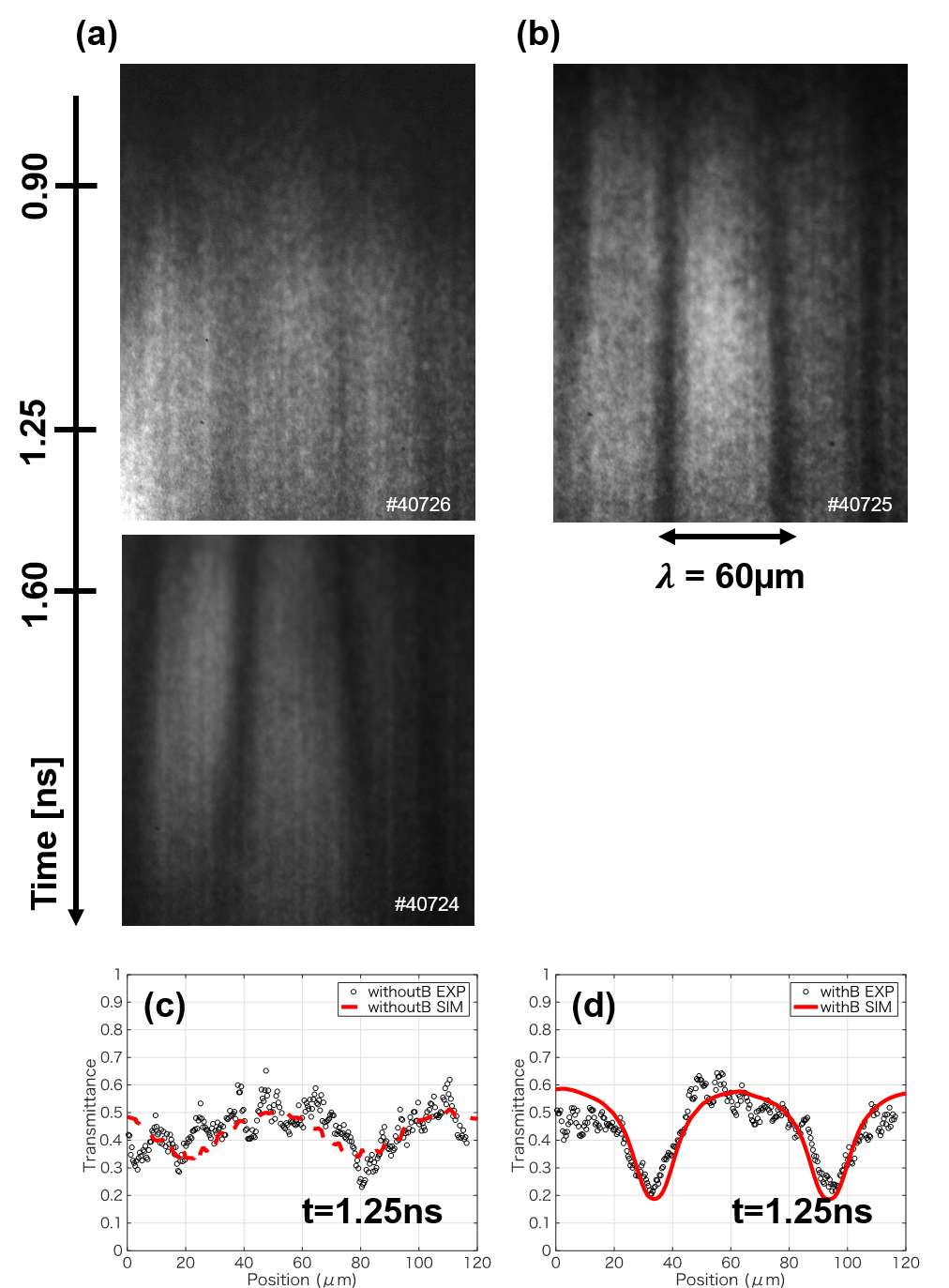}
\end{center}
\caption{(a),(b) Streak images of face-on backlight and (c),(d) profile at 1.25 ns, (a),(c) without a magnetic field and (b),(d) with a magnetic field. The time 1 ns is the drive laser peak time.} \label{fig:Figure2}
\end{figure}

Figure \ref{fig:Figure2} shows a face-on x-ray backlight image taken for a target with the modulation wavelength $\lambda = 60 \mu\mathrm{m}$. We defined the time of 1 ns ($t$ = 1 ns) to be the drive laser peak time.
Figures \ref{fig:Figure2}(a) and (b) show the cases without and with a magnetic field, respectively.
The RT growth is clearly visible: dark and light stripes, representing peaks and valleys of areal mass distribution, respectively, become more pronounced at the later regime.
Figures \ref{fig:Figure2}(c), (d) show the profiles of the x-ray transmittance at $t$ = 1.25 ns without a magnetic field and with a magnetic field, respectively.
The black dots are the experimental results of x-ray transmittance evaluated from the line-out of the face-on x-ray backlight image.
The red lines are the results of the two dimensional magnetohydrodynamic code PINOCO-2D-MHD\cite{Nagatomo2013} with the Braginskii model of electron thermal transport.
The modulations grow faster in an external magnetic field than in the absence of a magnetic field. 

\begin{figure}[ht]
 \centering 
 \includegraphics[width=80mm]{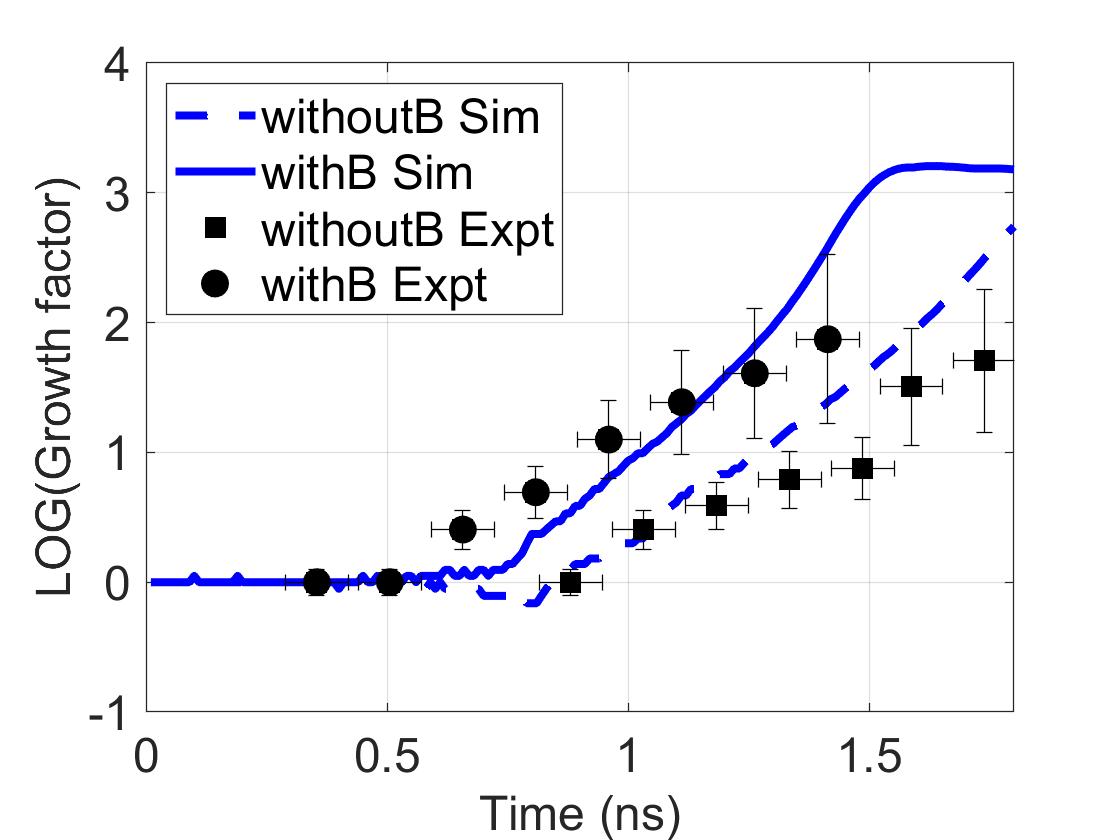}
\caption{Comparison of the experimentally measured growth factor and the predictions of the PINOCO-2D-MHD simulations for the 60-$\mu$m wavelength perturbations with an initial s of 1.0 $\pm$ 0.1 $\mu$m. The experimental data for the 1.2-ns gauss drive pulse are shown as dots and the calculated growth factor is shown as a solid line. Noise has been subtracted from the experimental data before being plotted with the simulation results. The 2-D hydrodynamic simulations are in good agreement with the experimental data.}
 \label{fig:Figure3}
\end{figure}

Figure \ref{fig:Figure3} shows the time evolution of the growth factors of the fundamental mode, where the growth factor is defined as the measured areal-density perturbation divided by the initial amplitude of 1.0 $\pm$ 0.1 $\mu$m.
The black dots are the experimental results and the blue curves are the results of the MHD simulation.
The simulation results exhibit a faster growth with a magnetic field than without a magnetic field.

\begin{figure}[ht]
	\centering 
	\includegraphics[width=85mm]{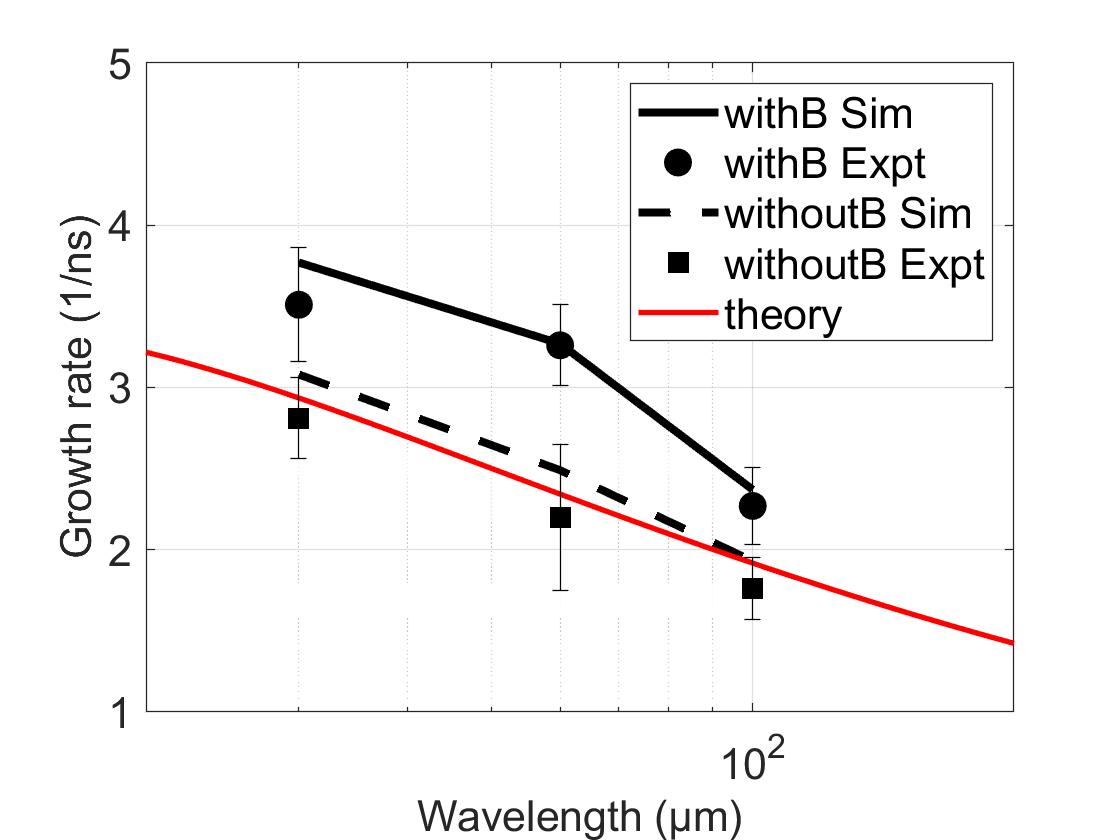}
   \caption{Growth rate of the RT instability obtained from the exponential fit to the data of the growth factor in a linear growth regime.}
	\label{fig:Figure4}
\end{figure}

Figure \ref{fig:Figure4} shows the growth rate of the ablative RT instability. The plotted values are summarized in Table \ref{table: growth_rate_summary}.
The growth rate of the RT instability is obtained from the exponential fit to the time history of the growth factor in a linear growth regime. We have defined the linear regime as that with an amplitude smaller than 10$\%$ of the wavelength. Similar results were also obtained in the other facilities for the case without a magnetic field \cite{Glendinning1997, Smalyuk2008}. Obviously, the growth rate in the external magnetic field are faster than that without a magnetic field. The experimental data are in good agreement with the simulation results.

\begin{table}
	\caption{\label{table: growth_rate_summary} Compilation of growth rates from experimental data and data from the PINOCO-2D-MHD simulation.}
	\begin{tabular}{cccccc}
		\hline
		$\lambda$ & $\Gamma_{\mathrm{Expt}}$ & $\Gamma_{\mathrm{Sim}}$ & $\Gamma_{\mathrm{Expt}}$ & $\Gamma_{\mathrm{Sim}}$\\
		\textbf{} & withoutB & withoutB & withB & withB\\
		 ($\mu$m) &(ns$^{-1}$) & (ns$^{-1}$) & (ns$^{-1}$) & (ns$^{-1}$) \\\hline \hline
		30 & 2.81 $\pm$ 0.25 & 3.08 & 3.51 $\pm$ 0.35 & 3.77 \\
		60 & 2.20 $\pm$ 0.45 & 2.49 & 3.26 $\pm$ 0.25 & 3.27 \\
		100 & 1.76 $\pm$ 0.19 & 1.93 & 2.27 $\pm$ 0.24 & 2.37 \\
	    \hline
	\end{tabular}
\end{table}


The growth rate $\gamma$ of the RT instability including the ablation effect was suggested by Bodner\cite{Bodner1974}, and is approximated by the modified Takabe formula\cite{Takabe1983}:
\begin{equation}
    \Gamma = \alpha \sqrt{\frac{kg}{1+kL}} - \beta k V_a, \label{eq:RT_growth}
\end{equation}
where $\alpha$ and $\beta$ are coefficients, $k$ is the wave number of the perturbation, $g$ is gravity, $L$ is the density scale length at the ablation surface, and $V_a$ is the ablation velocity.
In our experiment, $g = 75 \pm 5 \mathrm{\mu m/ns^2}$ is calculated from a trace of the laser-driven polystyrene foil observed using an x-ray streak camera and side-on backlighting. The results of the PINOCO-2D-MHD simulation reproduce the velocity of the accelerated polystyrene foil.
Near the laser peak time, $L \approx 1 \mathrm{\mu m}$ and $V_a \approx 2 \mathrm{\mu m/ns}$ are obtained from the simulation. Betti \textit{et al.} found an analytical solution\cite{Betti1998} for a plastic target, which corresponds to $\alpha$ = 0.98 and $\beta$ = 1.7 when approximated by Eq.\ref{eq:RT_growth}.
The theoretical growth rate is also shown in Fig.\ref{fig:Figure4} by the red curve, which confirm the accuracy of the experimental measurements and the MHD simulations.

\begin{figure}[ht]
	\centering 
	\includegraphics[width=70mm]{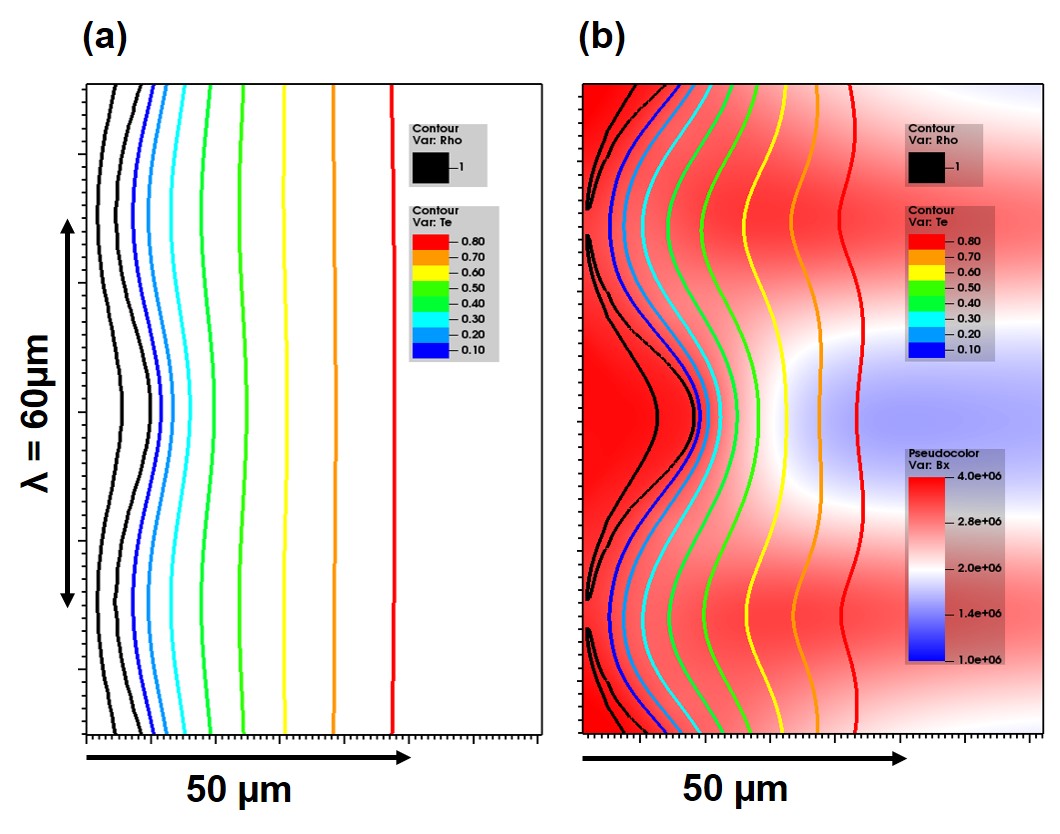}
   \caption{Ablation-front modulation creates stronger temperature gradients at the perturbation peaks and weaker gradients at the valleys. The heat flux is proportional to the gradients, which leads to a slightly enhanced heat flux at peaks and a reduced heat flux at valleys. The direction of the ablated plasma flow is normal to the target surface, and ablated plasma accumulates at the valleys of the sinusoidal perturbation. Therefore, the external magnetic field is compressed at the valleys and decompressed at the peaks of the sinusoidal perturbation.}
	\label{fig:Figure5}
\end{figure}

As a result of the perturbation growth, the peaks of the ablation-front ripple protrude into the hotter plasma corona, and the valleys recede toward the colder plasma corona.
Since the temperature perturbation is flat in the laser absorption region, the temperature gradients and heat fluxes are enhanced at the peaks and reduced at the valleys, as shown in Fig.\ref{fig:Figure5}(a). 
An excess in the heat flux speeds up the ablation front, while a deficiency slows the front down, a process called ablative stabilization. The ablative stabilization is more pronounced for shorter wavelengths because the temperature perturbation is localized in the vicinity of the unstable surface with a spatial extent of the order of the perturbation wavelength.

The hydrodynamic perturbation growth is affected by the external magnetic field as a result of the anisotropic thermal conductivity in the ablated plasma.
The magnetic field lines move together with the ablated plasma due to its large magnetic Reynolds number. 
The direction of the ablated plasma flow is normal to the target surface, and ablated plasma accumulates at the valleys of the sinusoidal perturbation. Therefore, the external magnetic field is compressed at the valleys and decompressed at the peaks of the sinusoidal perturbation, as shown in Fig.\ref{fig:Figure5}(b). The temperature increases at the valleys due to anisotropic thermal conduction in the perturbed magnetic field structure. The pressure distribution becomes spatially nonuniform.
The ablative stabilization is also reduced by less thermal smoothing of the temperature perturbation in the compressed magnetic field.
These lead to enhancement of the perturbation growth.

When the magnetic field causes the time scale of the thermal diffusivity to be longer than the hydrodynamic time scale, heat is trapped inside the ablation plasma, affecting the fluid motion.
The thermal diffusivity ($\eta$) is the thermal conductivity divided by the density and specific heat capacity at constant pressure.
The time scale of the thermal diffusivity in the compressed magnetic field ($\eta_{\bot}$) is locally smaller than the hydrodynamic time scale on the wavelength scale ($\lambda$) of the perturbation. This leads to the following inequalities:
\begin{equation}
    \frac{\lambda^2}{\eta}<\frac{\lambda}{C_s}<\frac{\lambda^2}{\eta_{\bot}}.
\end{equation}
Solving these inequalities for the wavelength, we obtain the condition that the magnetic field can affect the ablative RT instability as
\begin{equation}
    \frac{\eta}{C_s}\frac{1}{\sqrt{1+(\omega_\textrm{c} \tau_\textrm{ei})^2}}<\lambda<\frac{\eta}{C_s}. \label{eq:condition}
\end{equation}
For a 300-eV polystyrene plasma at critical density for 351-nm laser beams, $\frac{\eta}{C_s} \approx 120 \mu m$.
In the compressed magnetic field, the Hall parameter is about 1 $\sim$ 6. Therefore, the growth rates of the ablative RT instability in our experiment with $\lambda$ = 30-100 $\mu$m are enhanced by the reduction of thermal transport due to the magnetic field.

In summary, RT instability growth experiments were performed to demonstrate unavoidable perturbation growth due to a magnetic field. The unstable modulation growth in an external magnetic field was strongly increased compared to the growth in the absence of a field. The experiment was reproduced by a magnetohydrodynamic simulation based on a Braginskii model of electron thermal transport.

The external magnetic field reduces the electron thermal conduction across the magnetic field lines because the Larmor radius of the thermal electrons in the external magnetic field is much shorter than the temperature scale length. The thermal conduction suppression leads to spatially nonuniform pressure and reduced thermal ablative stabilization, which in turn increases the growth of ablative RT instability. 
We also obtain the condition that the magnetic field can affect the ablative RT instability as Eq.\ref{eq:condition}.
The stronger the magnetic field, the higher the modes of RT growth affected. In other words, by reducing the thermal transport at the instability growth front, the magnetization allows higher modes of RT instability to grow. 

These effects must be considered in the design of magnetically assisted ICF, which may be an alternative to fusion ignition schemes. Srinivasan \textit{et al.} have pointed out that the strengths of the self-generated magnetic field and the Hall parameter in National Ignition Facility implosions are estimated to be of the order of $10^2$--$10^3$ T and in the range between 0.1 and 1, respectively \cite{Srinivasan2012}. In such a strong self-generated magnetic field, anisotropic thermal conduction may affect hydrodynamic growth.
The thermal conduction suppression due to a magnetic field at ignition-scale lasers requires further investigation.

\section*{acknowledgments}
The authors wish to thank the technical support staff of ILE and the Cyber Media Center at Osaka University for assistance with the laser operation, target fabrication, plasma diagnostics, and computer simulations. 
This research used the computational resources of the HPCI system provided by Information Technology Center, Nagoya University through the HPCI System Research Project (Project ID: hp180093).
This work was supported by the Collaboration Research Program between the National Institute for Fusion Science and the Institute of Laser Engineering at Osaka University, and by the Japanese Ministry of Education, Science, Sports, and Culture through Grants-in-Aid, KAKENHI (Grants No. 24684044, 25630419, 26287147, 15K17798, 15K21767, 15KK0163, 16K13918, 16H02245, and 17K05728), Bilateral Program for Supporting International Joint Research by JSPS, Grants-in-Aid for Fellows by Japan Society for The Promotion of Science (Grant No. 14J06592, 15J00850, 15J00902, 15J02622, 17J07212, 18J01627, 18J11119, and 18J11354), Matsuo Research Foundation, and the Research Foundation for Opto-Science and Technology.

\bibliographystyle{h-physrev3}

\end{document}